\documentclass[twoside,slac_one]{revtex4}
\usepackage{graphicx}
\usepackage{fancyhdr}
\usepackage{amsmath} 
\usepackage{bm}
\usepackage{amsxtra}
\usepackage{amssymb}
\usepackage{amsthm}
\usepackage{latexsym}
\usepackage{lscape}

\def\simge{\mathrel{%
   \rlap{\raise 0.511ex \hbox{$>$}}{\lower 0.511ex \hbox{$\sim$}}}}
\def\simle{\mathrel{
   \rlap{\raise 0.511ex \hbox{$<$}}{\lower 0.511ex \hbox{$\sim$}}}}

\newcommand{\MET}{\mbox{$E\kern-0.60em\raise0.10ex\hbox{/}_{T}$}}

\begin{document}

\title{A Fast Hardware Tracker for the ATLAS Trigger System}

%

\author{M.S. Neubauer on behalf of the ATLAS Collaboration}
\affiliation{Department of Physics, University of Illinois at
  Urbana-Champaign, Urbana, IL, USA}

\begin{abstract}
In hadron collider experiments, triggering the detector to store
interesting events for offline analysis is a challenge due to the high
rates and multiplicities of particles produced. The LHC will soon
operate at a center-of-mass energy of 14 TeV and at high instantaneous
luminosities of the order of $10^{34}$ to $10^{35}$ cm$^{-2}$ s$^{-1}$. A
multi-level trigger strategy is used in ATLAS, with the first level
(LVL1) implemented in hardware and the second and third levels (LVL2 and
EF) implemented in a large computer farm. Maintaining high trigger
efficiency for the physics we are most interested in while at the same
time suppressing high rate physics from inclusive QCD processes is a
difficult but important problem. It is essential that the trigger
system be flexible and robust, with sufficient redundancy and
operating margin. Providing high quality track reconstruction over the
full ATLAS detector by the start of processing at LVL2 is an important
element to achieve these needs. As the instantaneous luminosity
increases, the computational load on the LVL2 system will significantly
increase due to the need for more sophisticated algorithms to suppress
backgrounds.

\parindent=10pt
The Fast Tracker (FTK) is a proposed upgrade to the ATLAS trigger
system. It is designed to enable early rejection of background
events and thus leave more LVL2 execution time by moving track
reconstruction into a hardware system that takes massively parallel
processing to the extreme. The FTK system completes global track
reconstruction with near offline resolution shortly after the start of
LVL2 processing by rapidly finding and fitting tracks in the inner
detector for events passing LVL1 using pattern recognition from a large,
pre-computed bank of possible hit patterns. We describe the FTK system
design and expected performance in the areas of $b$-tagging,
$\tau$-tagging, and lepton isolation which play and important role in the
ATLAS physics program.
\end{abstract}

\maketitle

\thispagestyle{fancy}

\section{Introduction}
The Large Hadron Collider will collide protons with planned bunch
spacing of 25 nanoseconds and center-of-mass energy of 14 TeV. At an
instantanous luminosity of $3 \times 10^{34}$ cm$^{-2}$ s$^{-1}$, each
collision on average produces 75 minimum-bias interactions that result
in high detector occupancy and create a challenging environment for
event readout and reconstruction. On one hand, limited data store
bandwidth demands a significant online rate reduction of 5-6 orders of
magnitude. On the other hand, events with interesting physics
signatures must be selected very efficiently from the vast LHC background.

\section{Current ATLAS Trigger System}

The primary role of the trigger system in the ATLAS experiment is to
search for interesting but extremely rare processes in proton
collisions at the LHC that are hidden in much larger background levels
and initiate an archive of detector data for such processes. Only a
tiny fraction of the produced collisions can be stored on tape and an
enormous real-time data reduction is needed. This requires massive
computational power to minimize the online execution time of complex
algorithms. A multi-level trigger is an effective solution for an
otherwise intractable problem.

The rate of recorded events is limited to about 300 Hz due to
technology and resource limitations. Since the proton-proton
interaction rate at design luminosity is 1 GHz, the trigger must
maintain an overall rejection factor of 5 million against minimum-bias
processes while retaining maximum efficiency for the new physics
sought by ATLAS. The Level-1 (LVL1) trigger system uses a subset of
the total detector information to make a decision in less than 2.5
microseconds regarding whether or not to continue processing an event,
reducing the data rate to approximately 75 kHz (limited by the
detector readout system, which is upgradeable to 100 kHz). The Level-2
(LVL2) and Event Filter (EF) together form the High-Level Trigger
(HLT). They provide the reduction to a final data-taking rate of
approximately 300 Hz. The LVL1 trigger is implemented using custom
electronics, while the HLT is almost entirely based on commercially
available computers and networking hardware.

The LVL1 trigger searches for signatures from high pT  muons,
electrons/photons, jets, and τ-leptons decaying into hadrons. It also
selects topologies with large missing transverse energy ($\MET$) and
large total transverse energy $E_T$. The LVL1 trigger uses
reduced-granularity information from a subset of detectors: the
Resistive Plate Chambers (RPC) and Thin-Gap Chambers (TGC) for high pT
muons, and all the calorimeter sub-systems for electromagnetic
clusters, jets, $\tau$-leptons, $\MET$, and large $E_T$. The LVL1
decision time must be less than 2.5 microseconds given the maximum
readout rate of 75 KHz.

Although adding tracking information at this stage would be extremely
beneficial to discern certain physics objects, it is not currently
possible given the small timing window (2.5 $\mu s$) available to
LVL1. Moreover, even projected CPU farms that  constitute the LVL2
trigger cannot perform global track reconstruction within their time
budget of approximately 10 ms. Instead, at LVL2 does limited tracking
inside Regions of Interest (ROI) identified by the LVL1 trigger. These
are regions of the detector where the LVL1 trigger has identified LVL1
trigger objects of interest within the event.

The LVL2 trigger uses ROI information on coordinates, energy, and type
of signatures to limit the amount of data that must be transferred
from the detector readout. It uses dedicated algorithms to analyze
full-granularity and full-precision data from all detectors. The LVL2
trigger reduces the event rate to below 3.5 kHz, with an average event
processing time of approximately 10 milliseconds. The EF uses offline
analysis procedures on complete events to further select events down
to a rate that can be recorded for subsequent offline analysis. It
reduces the event rate to approximately 300 Hz, with an average event
processing time of order one second.

\section{Why Add a Hardware Tracker?}

An important lesson from past experience at high-luminosity hadron
collider experiments is that controlling trigger rates can be
extremely challenging as the luminosity increases, physics goals
change in response to new discoveries, and the detector ages.  It is
thus essential that the trigger system be flexible and robust, with
sufficient redundancy and operating margin.  Providing high quality
track reconstruction over the full ATLAS detector by the start of
processing in the LVL2 computer farm is an important element in
achieving these goals. This has certainly been the case in CDF Run II,
where the availability of near offline-quality tracks at the trigger
hardware level using chamber and silicon hit information has
significantly extended the physics capabilities of the experiment
(e.g. triggering on displaced tracks which are rich in $b$-hadron decays). 

As the instantaneous luminosity increases, the trigger task becomes
more difficult since the signal rates grow linearly but the background
rates grow more quickly. At $3 \times 10^{34}$ cm$^{-2}$ s$^{-1}$,
each bunch crossing contains ~75 pile-up interactions (on average)
that need to be distinguished by the trigger from a hard scattering
processing of interest. The computational load on the LVL2 system will
significantly increase due to both the need for more sophisticated
algorithms to suppress backgrounds and more ROIs to process. Until the
new physics processes that will be seen at the LHC are known, it is not
obvious how to best deal with this challenge since we do not know the
momentum scale of the trigger objects that need to be considered or
what mix of lepton, jets, and $\MET$ in the trigger menu will be
optimal. Increasing the amount of information available at the start
of LVL2 processing through early track reconstruction will enhance the
trigger flexibility needed during high luminosity LHC running.
  
We can reasonably expect third-generation fermions to play an
important role in LHC physics given central role that mass plays in
electroweak symmetry breaking. An example of the utility of early
track reconstruction is the selection of events containing
third-generation fermions, either a $b$-quark or a $\tau$-lepton. For
$b$-quarks, the LVL1 trigger selects a generic jet. For hadronic
$tau$-lepton decays, the LVL1 trigger selects a narrow jet. The
challenge comes from the enormous background from QCD produced light
quark and gluon jets, which can be suppressed using tracking
information. Tracks coming from a secondary vertex or not pointing to
beamline identify $b$-quark jets, while $\tau$ jets (from hardronic
$\tau$ decays).

While we do not yet know what Nature has in store for the LHC, we do
know that efficient triggering on final states involving electrons, muons,
missing transverse energy (MET), and/or $3^{\rm{rd}}$-generation
fermions will be crucial for achieving the physics goals of the ATLAS
experiment. Bottom quarks and $\tau$-leptons produce very specific
tracking signatures which can be exploited in triggering. At high
luminosity, sophisticated algorithms run at LVL2 will be needed to
suppress backgrounds. In the current system, the LVL2 farm is
burdened with reconstructing tracks with the ROIs. Also, global
tracking will be important for maintaining reasonable trigger rates
and signal efficiency at high interaction pile-up (large
luminosity). In particular, the use of primary vertexing and
subsequent charged lepton association for improved isolation will
improve hadronic jet background rejection. With global tracking, we
can also use $b$-decay events with final state hadrons identified
outside of ROIs and construct additional combined triggers
(e.g. lepton + track). Finally, it is possible to construct
track-based MET and jet objects, possibly combined with calorimeter
information, that utilize primary vertexing to maintain stable and
efficient triggering under high luminosity running conditions.

\section{The FTK System Description}

We describe a system to enable early rejection of background events
and more LVL2 execution time for sophisticated algorithms by moving
track reconstruction into a hardware system
(FTK)~\cite{ieee10,alb01,alb02} with massively-parallel processing 
that produces global track reconstruction with nearly offline
resolution close to the start of LVL2 processing. The FTK system is
inspired by the success of the Silicon Vertex Trigger (SVT)~\cite{svt}
in the CDF detector. FTK operates in parallel with the normal silicon
detector readout following each LVL1 trigger and reconstructs
tracks over the entire detector volume (up to $|\eta|$ of 2.5) in
under a millisecond. Having tracks available by the beginning of
LVL2 processing allows reduced LVL1 $p_{T}$ thresholds since the
LVL2 trigger is able to reject non-interesting events more
quickly. Furthermore, since the LVL2 system is freed of it heavy
tracking load, the extra processing time becomes available for more
advanced online algorithms.

\subsection{Segmentation and Data Flow}

\begin{figure}[h]
\centering
\includegraphics[width=5in]{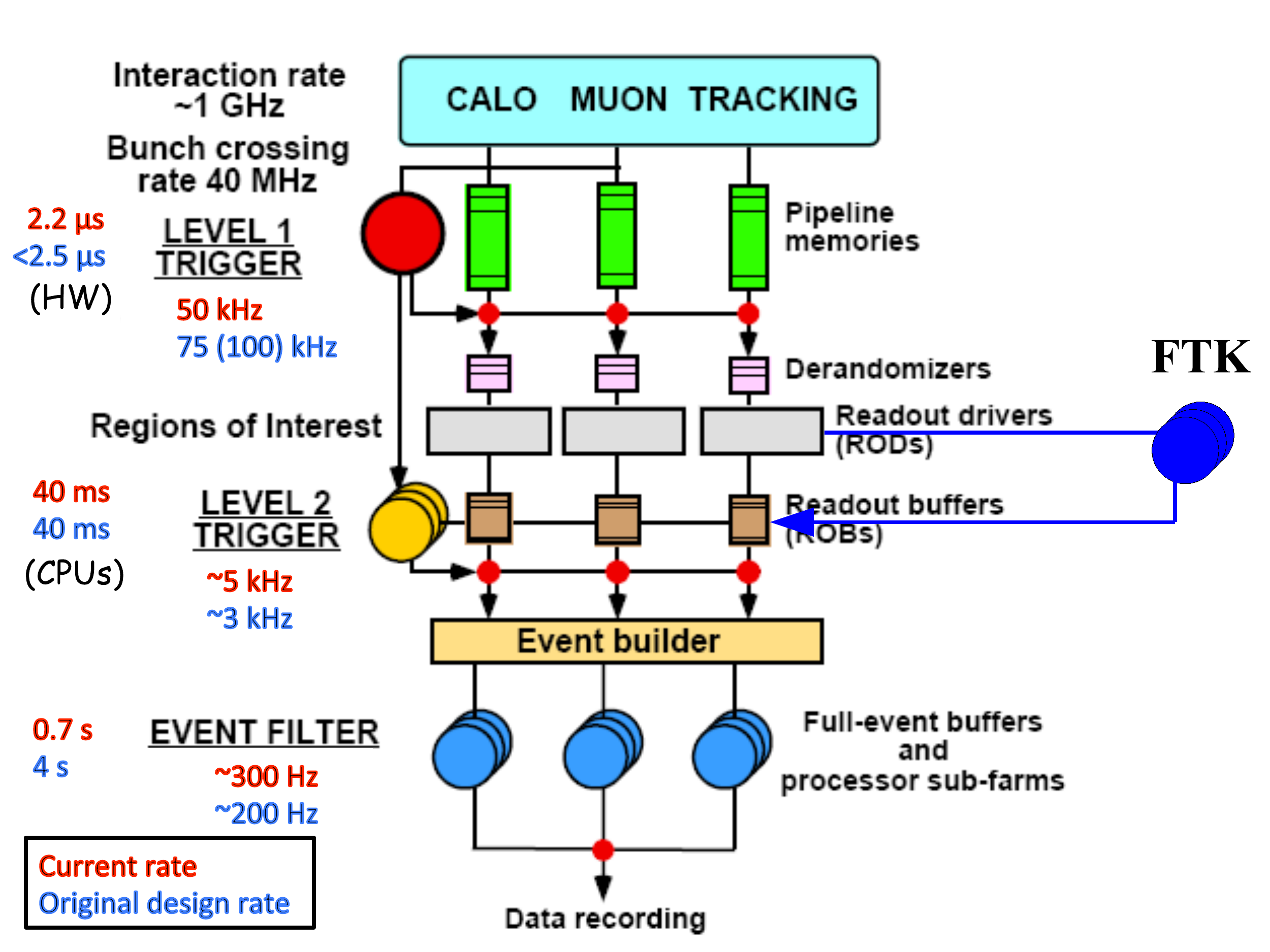}
\caption{ATLAS trigger system and its integration with FTK.} 
\label{trigger_system}
\end{figure}

In order to perform tracking with good efficiency and resolution, FTK
uses data from the two inner-most subsystems of the ATLAS Inner
Detector~\cite{atlas}. Pixel sensors provide a two-dimensional
measurement of hit position and include over 80 million readout
channels. SCT layers are arranged in pairs of axial and narrow-angle
stereo strips and consist of 6 million channels. A typical track
passes through 11 detector layers and can be reconstructed from the 14
coordinate measurements ($3 \cdot 2$ from two-dimensional pixels and
$8$ from SCT). Custom designed splitters duplicate data output by the
SCT and Pixel Readout Drivers (RODs) and send it to FTK. Since FTK
non-invasively eavesdrops on the Inner Detector data, it easily
integrates with the current ATLAS trigger system
(Fig.~\ref{trigger_system}).

\begin{figure}[h]
\centering
\includegraphics[width=6in]{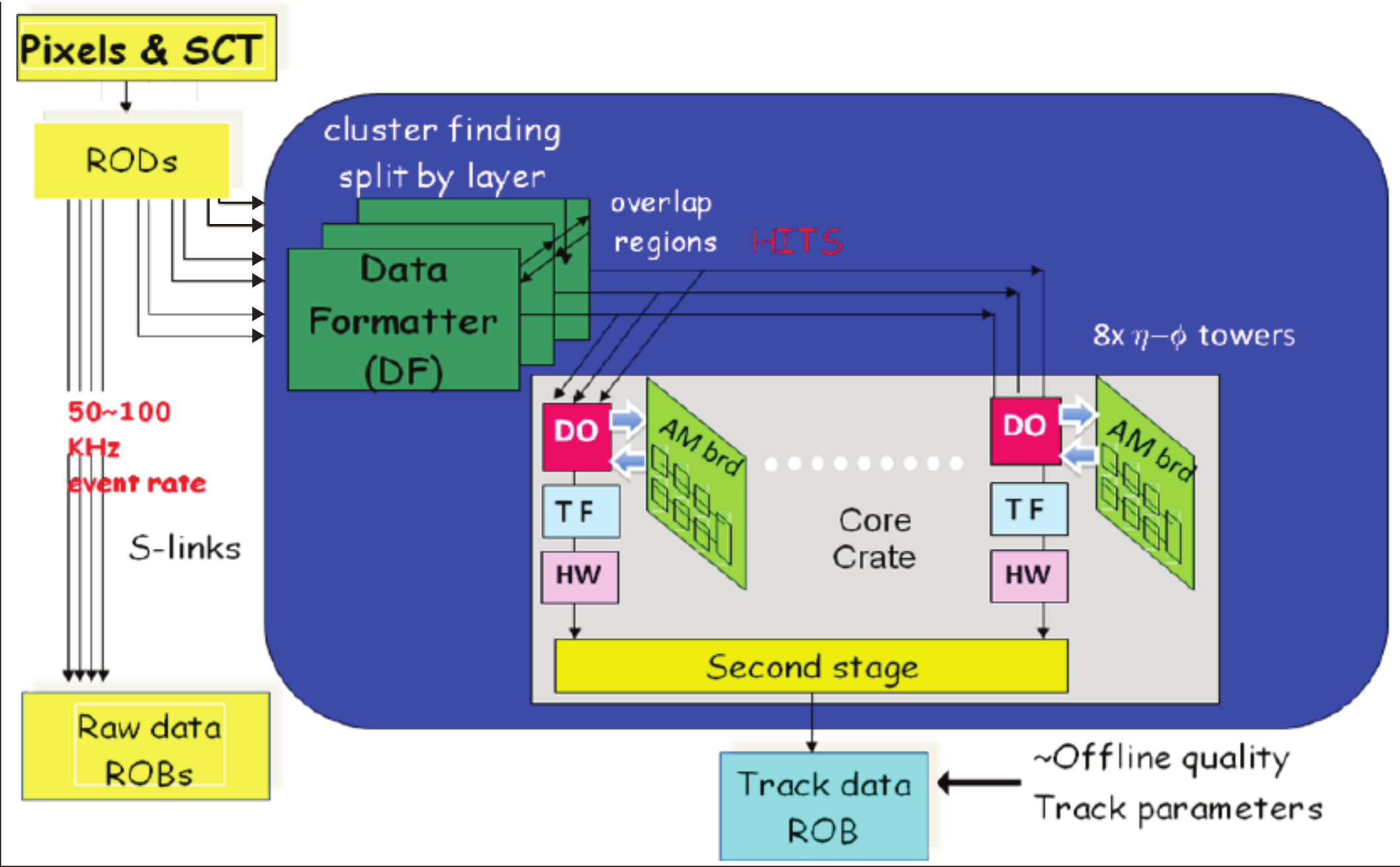}
\caption{Schematic diagram of the FTK data flow}
\label{data_flow}
\end{figure}

In order to increase the overall throughput of the system, FTK splits
incoming data into 8 or more {\it regions} in $\phi$ with sufficient
overlap to account for inefficiencies at the edges. Each region is
served by a separate crate that consists of the subsystems shown in
Fig.~\ref{data_flow}. Raw Pixel and SCT data are first received by the
Data Formatter that uses fast FPGA's to perform
clustering~\cite{clust}. Clustered hits proceed into one of six Data
Organizers depending on which of six $\eta-\phi$ towers the hits
reside, where they are buffered and merged into coarse superstrips 
to be used in pattern recognition. The superstrips are then sent into a
pipelined array of Associative Memory (AM) that perform fast pattern
finding using a pre-calculated table of particle trajectories. Matched
patterns are reconnected with their corresponding full-resolution hits
in their corresponding Data Organizer and sent to the Track
Fitters. After removal of duplicate tracks, with each phi-region a
Read-Out Buffer (ROB) is associated that receives the track data
(formatted as a ROD fragment). The ROBs transfer the output on request
to the LVL2 trigger.

\subsection{Pattern Recognition in Associative Memories}

\begin{figure}[h]
\centering
\includegraphics[width=5in]{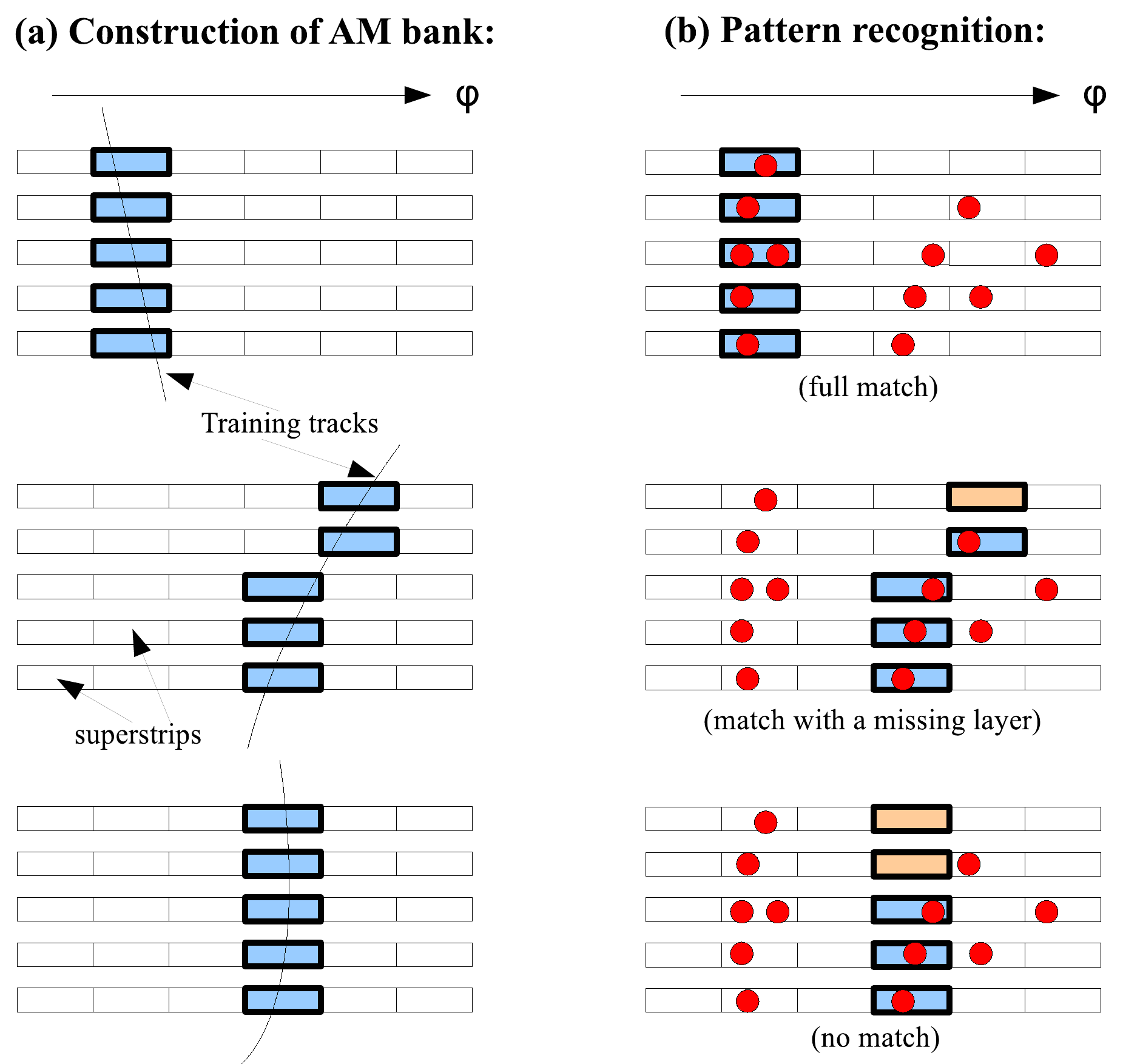}
\caption{Associative Memory and pattern bank operation} \label{am_bank}
\end{figure}

Luminosities above $10^{34}~cm^{-2}s^{-1}$ combined with 86 million
readout channels create a unique combinatorial challenge for
tracking. FTK overcomes this with the help of specialized AM hardware
that implements a massive, ultra-fast lookup table that enumerates all
realistic particle trajectories (patterns) through the 11 detector
layers~\cite{vlsi}. In order to keep the size of the trajectory lookup
under control, detector hits are merged into coarse-resolution {\it
  superstrips} having a width of a few millimeters\footnote{These
  coarse superstrips are only used in the pattern recognition stage;
  all final fits are performed with full resolution hits.}. The
pattern bank is precalculated either from single track Monte-Carlo or
from real data events (Fig.~\ref{am_bank}(a)) and stored in the AM.

Each pattern in the AM includes its own comparison logic. When hits
from a given event enter the AM, they are simultaneously compared
with millions of pre-stored patterns. In order to account for
inefficiencies in individual detector layers, FTK also matches
patterns with one missing layer (Fig.~\ref{am_bank}(b)). 

\subsection{Track Fitting}
FTK computes five track helix parameters (curvature, $d_{0}$ etc) and
a $\chi^2$ quality of fit from the {\it full resolution} hits within
each matched pattern. Since patterns are constructed from
reduced-granularity superstrips, multiple full-resolution hits can
belong to a given superstrip. This results in some ambiguity, which is
resolved by fitting all combinations within the superstrips.

Performing full $\chi^2$ minimization with respect to five parameters
is an extremely slow procedure. Instead, FTK reduces the track fitting
problem to a set of scalar products, which can be computed efficiently
using DSP units in modern commercial FPGA's. This is done by arranging
geometrically similar patterns into a number of groups (called {\it
  sectors}), so that within each sector the relationship between hit
positions ($x_{j}$) and track parameters ($p_{i}$) is approximately
linear: 
\begin{equation}
  p_{i} = \sum_{j=1}^{14} c_{ij} \cdot x_{j} + q_{i}
\label{eq-linear}
\end{equation}
The fitting coefficients for each sector are precomputed from the same
training data that was used in pattern generation. An added advantage
of this approach is that when real detector hits are used in training,
misalignments and other detector effects are automatically taken into
account.

\begin{figure}[h]
\centering
\includegraphics[width=3in]{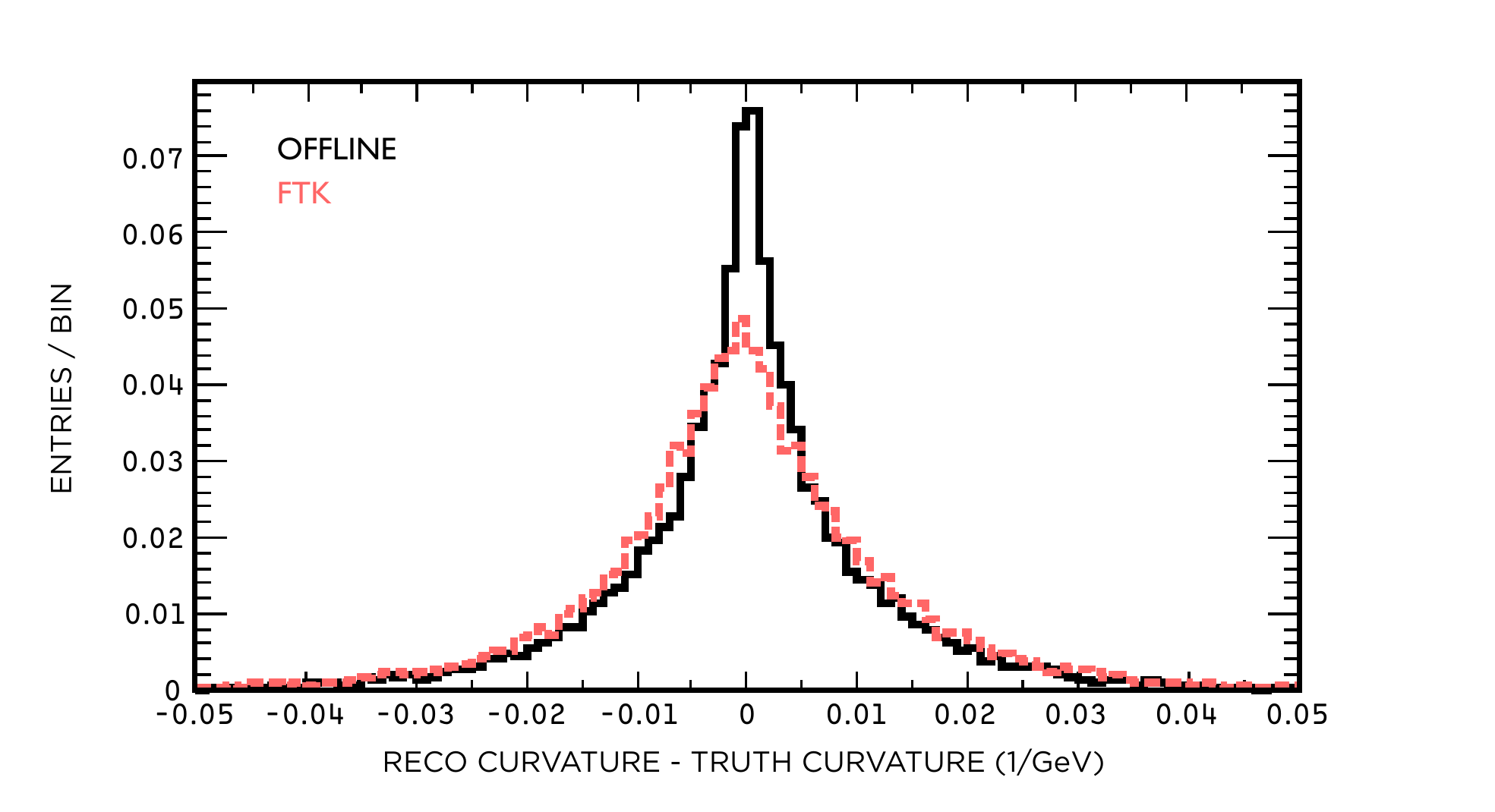}
\includegraphics[width=3in]{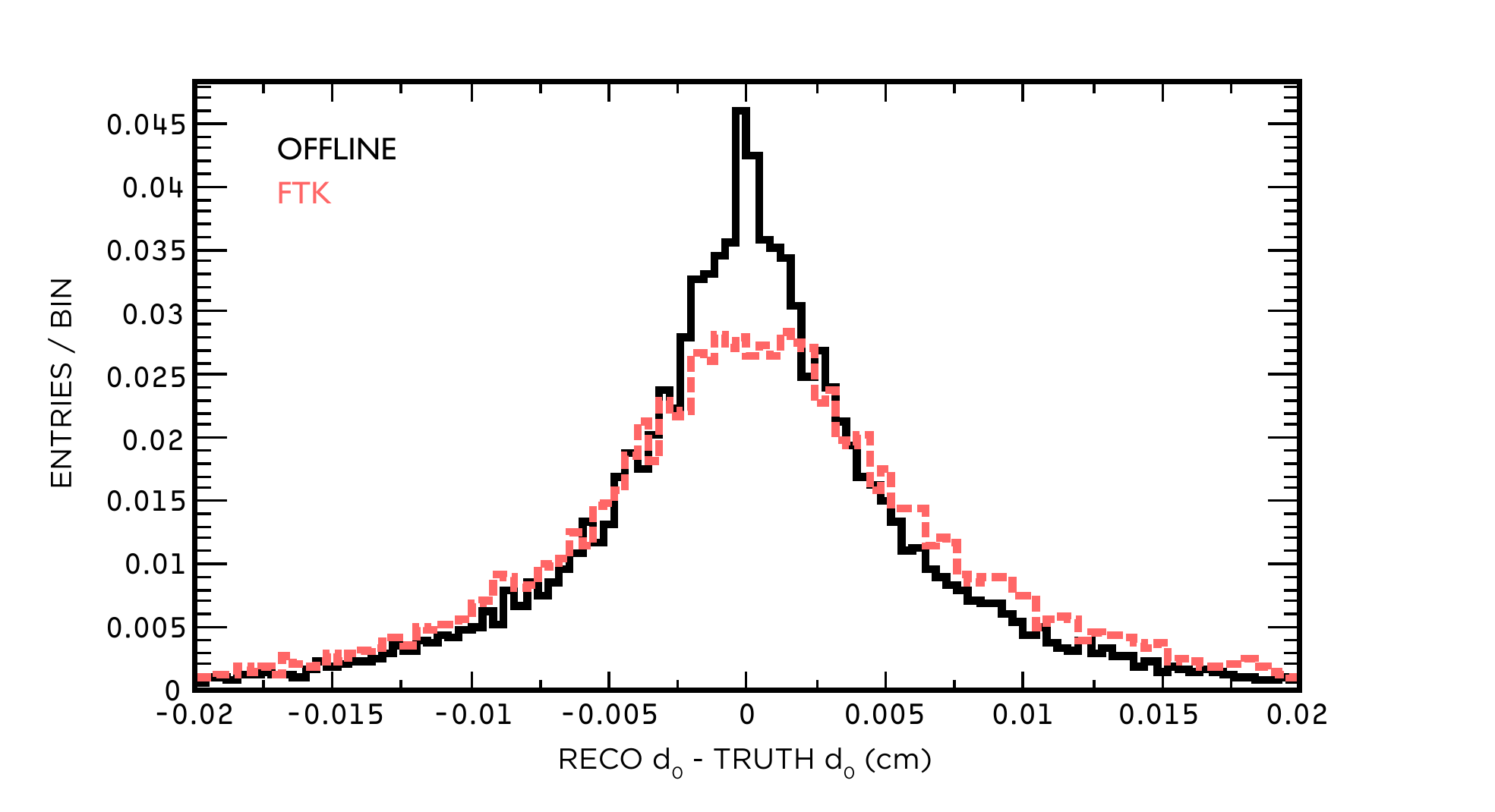}
\includegraphics[width=3in]{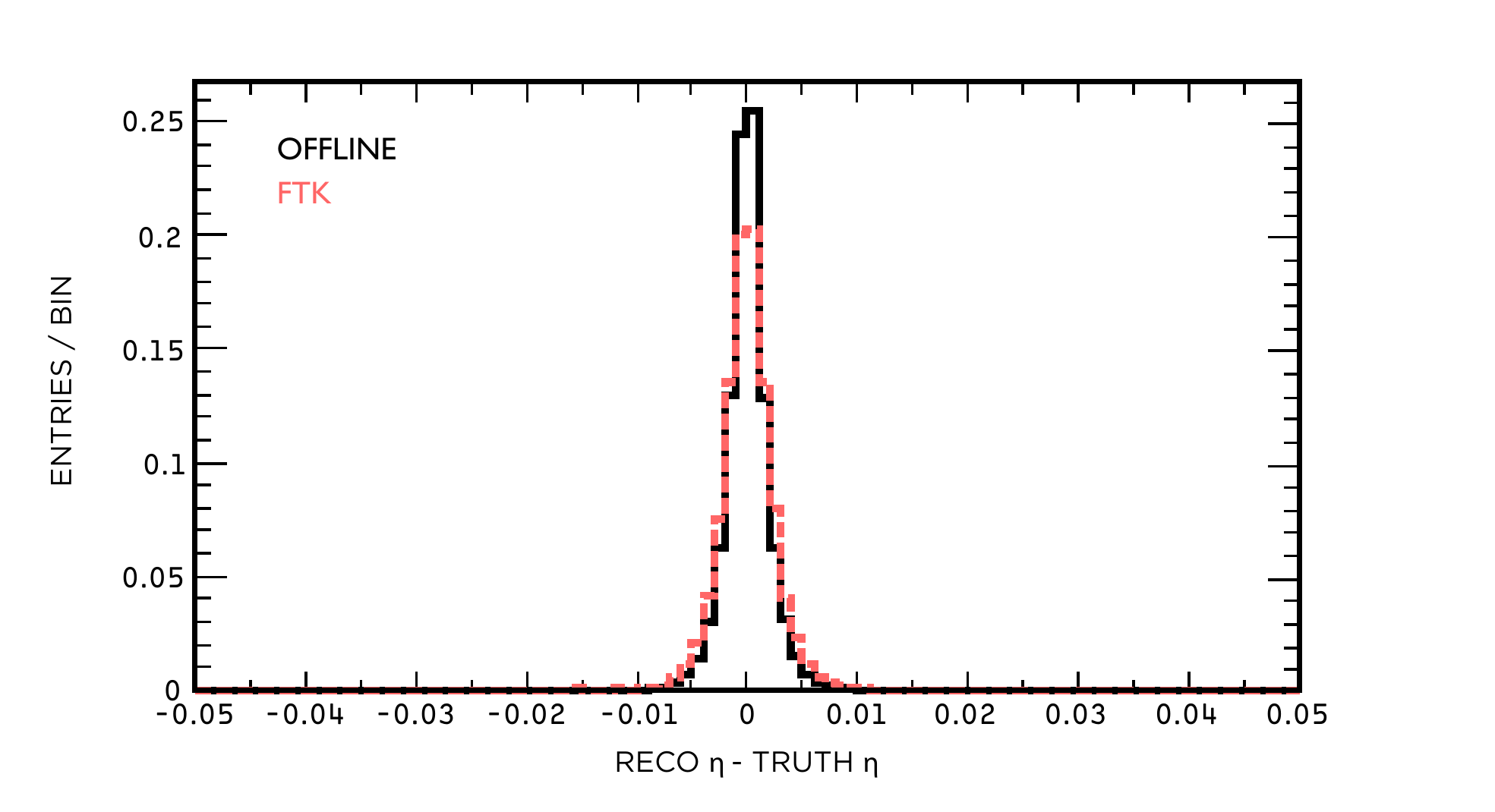}
\includegraphics[width=3in]{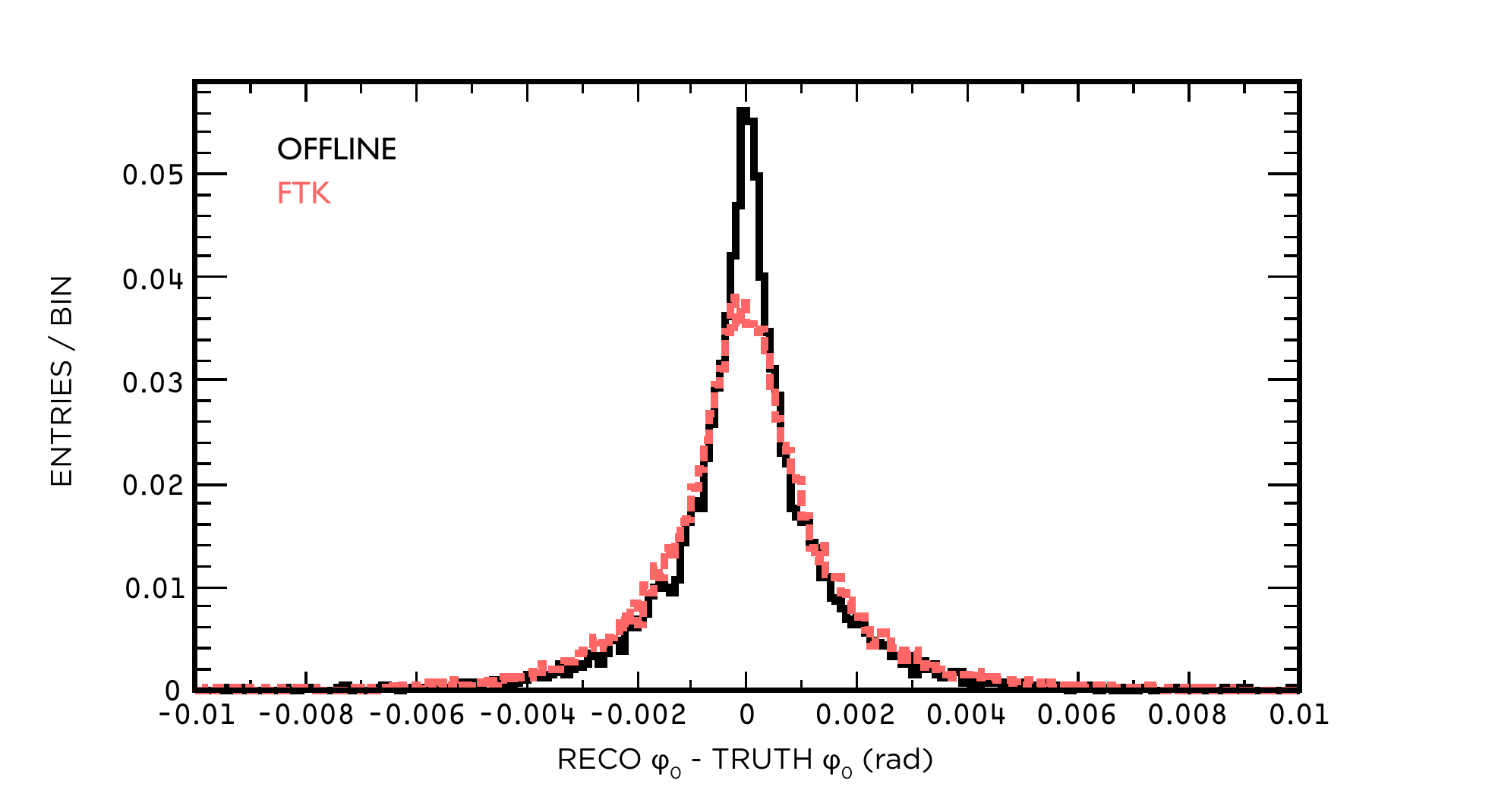}
\includegraphics[width=3in]{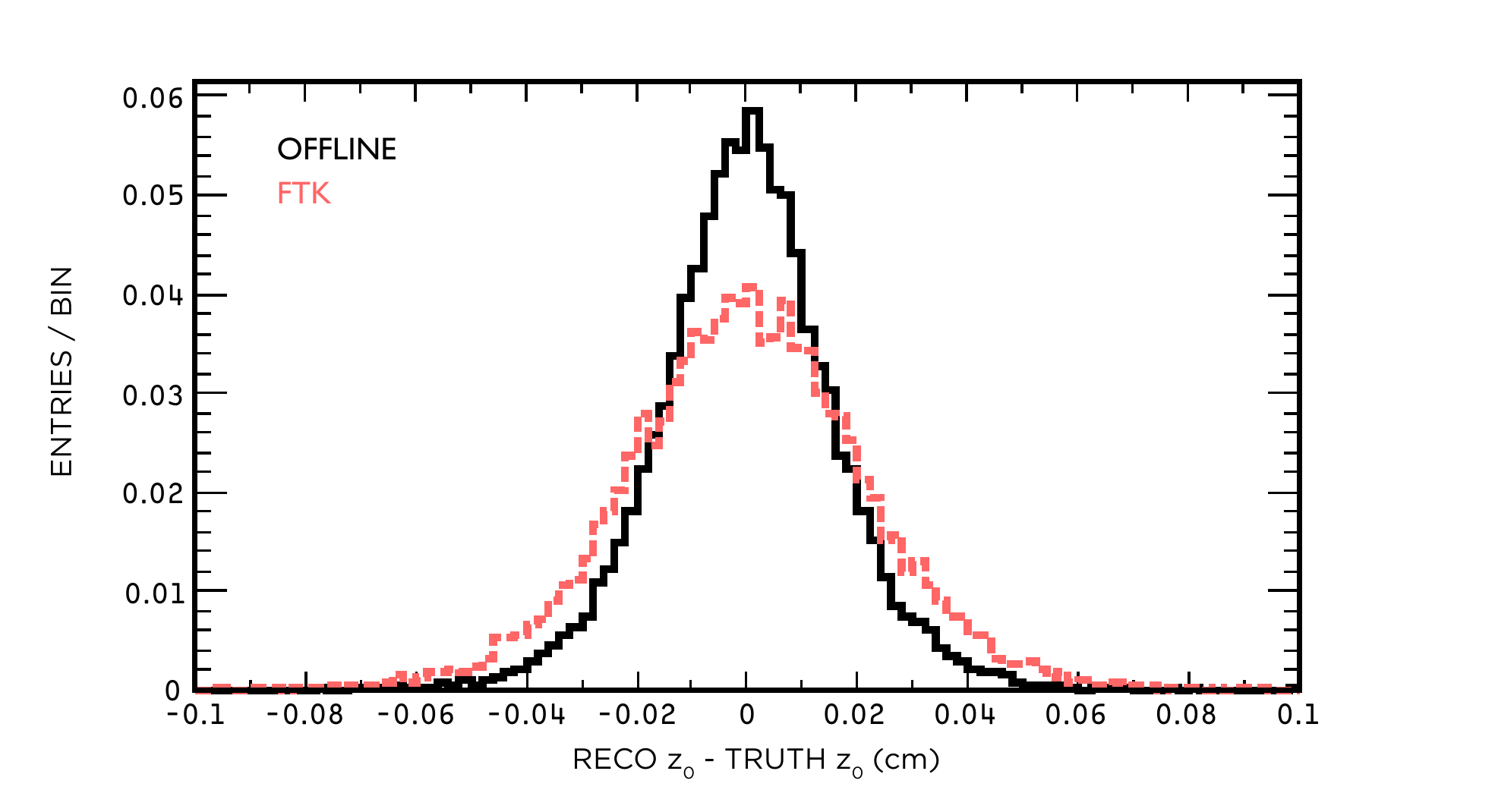}
\caption{Impact Parameter resolution for all primary tracks with
  $p_{T} > $ 1 GeV in $WH$ ($H\rightarrow b\bar{b}$) events at the design
  luminosity - comparison between FTK and offline
  tracking} \label{resol} 
\end{figure}

Overall, the linearized approach allows FTK to achieve near-offline
resolution as shown in Figure~\ref{resol}, with a fitting rate of
about 1 fit per nanosecond.

\section{Timing}

An important consideration the FTK design is the execution time for
typical events at high luminosity. The expected execution time of the
current ATLAS trigger system and the FTK system for $WH$ events
simulated at $3 \times 10^{34}$ cm$^{-2}$ s$^{-1}$ was evaluated. The
preliminary results are that while the current system requires
$\sim25$ msec per jet ROI (and there will be several of these per
event at an realistic jet $E_T$ threshold), global tracking of the
entire detector can be done by the FTK system in $\sim25~\mu$sec per
event for a track $p_T$ threshold of 1 GeV.

\section{Expected Performance}
\begin{figure}[h]
\centering
\includegraphics[width=80mm]{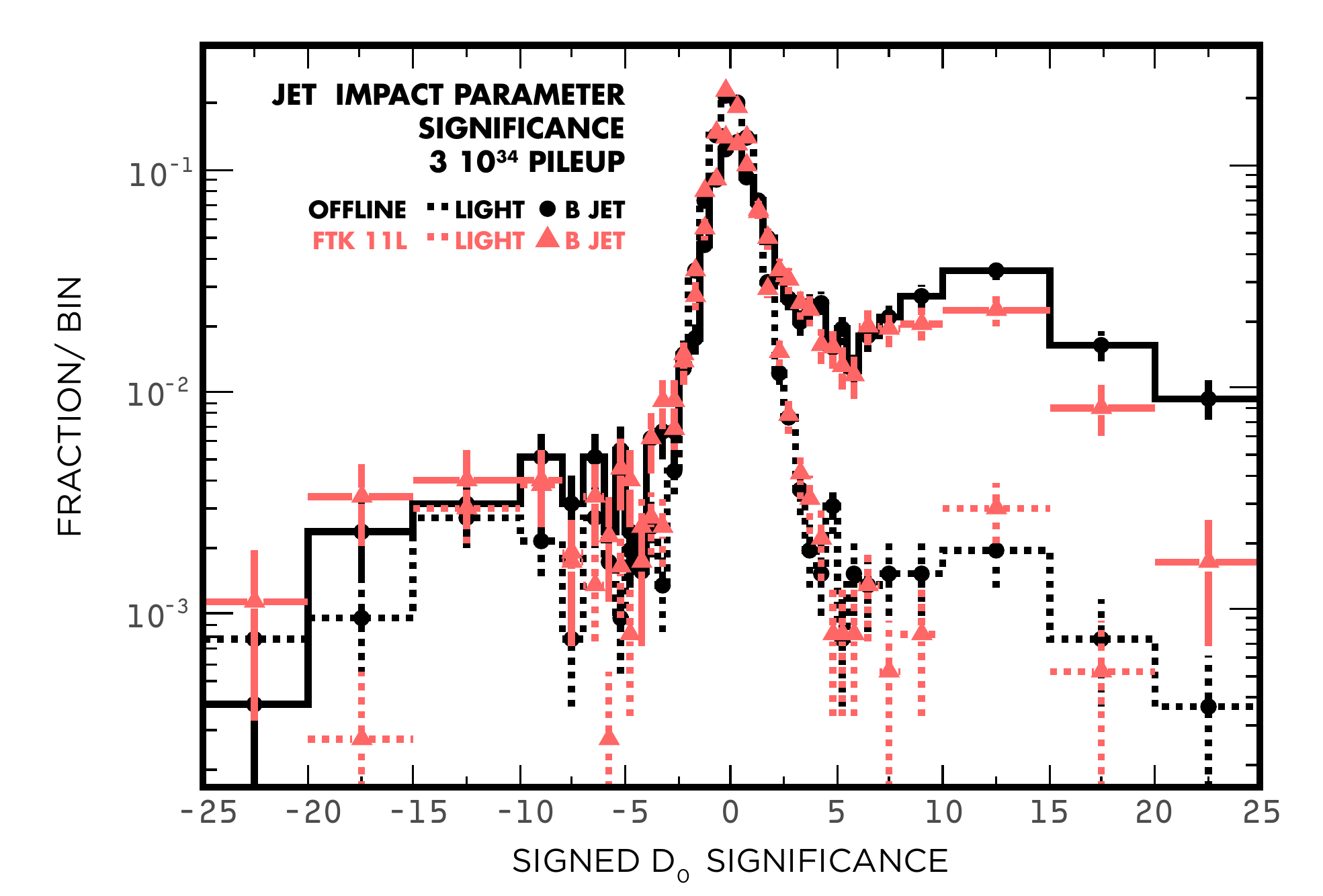}
\caption{Comparison of the signed impact parameter significance for
  $WH$ events simulated at $3 \times 10^{34}$ cm$^{-2}$ s$^{-1}$
  between offline tracks and FTK tracks.
}
\label{perf_btag}
\end{figure}

\begin{figure}[h]
\centering
\includegraphics[width=84mm]{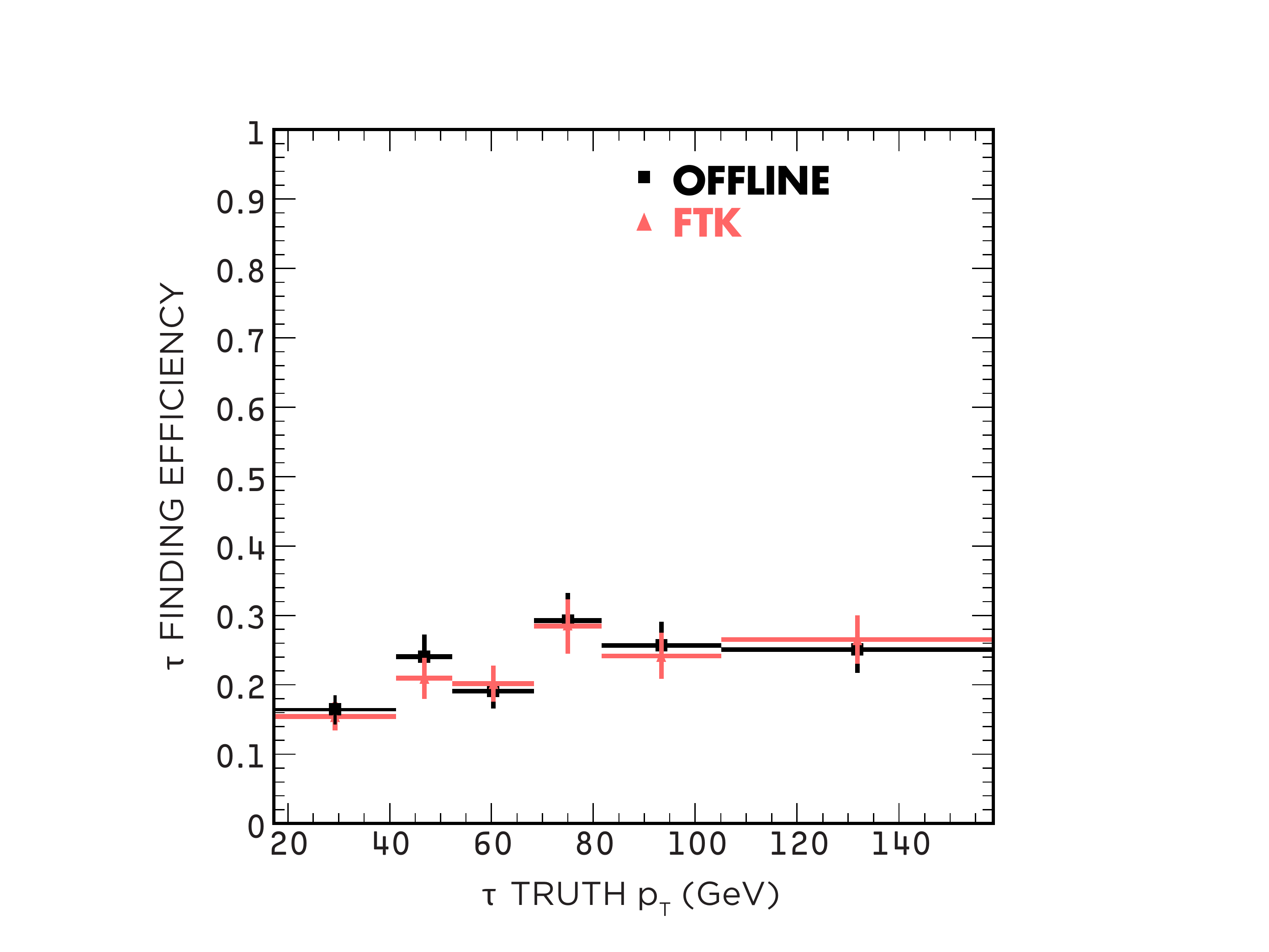}
\includegraphics[width=84mm]{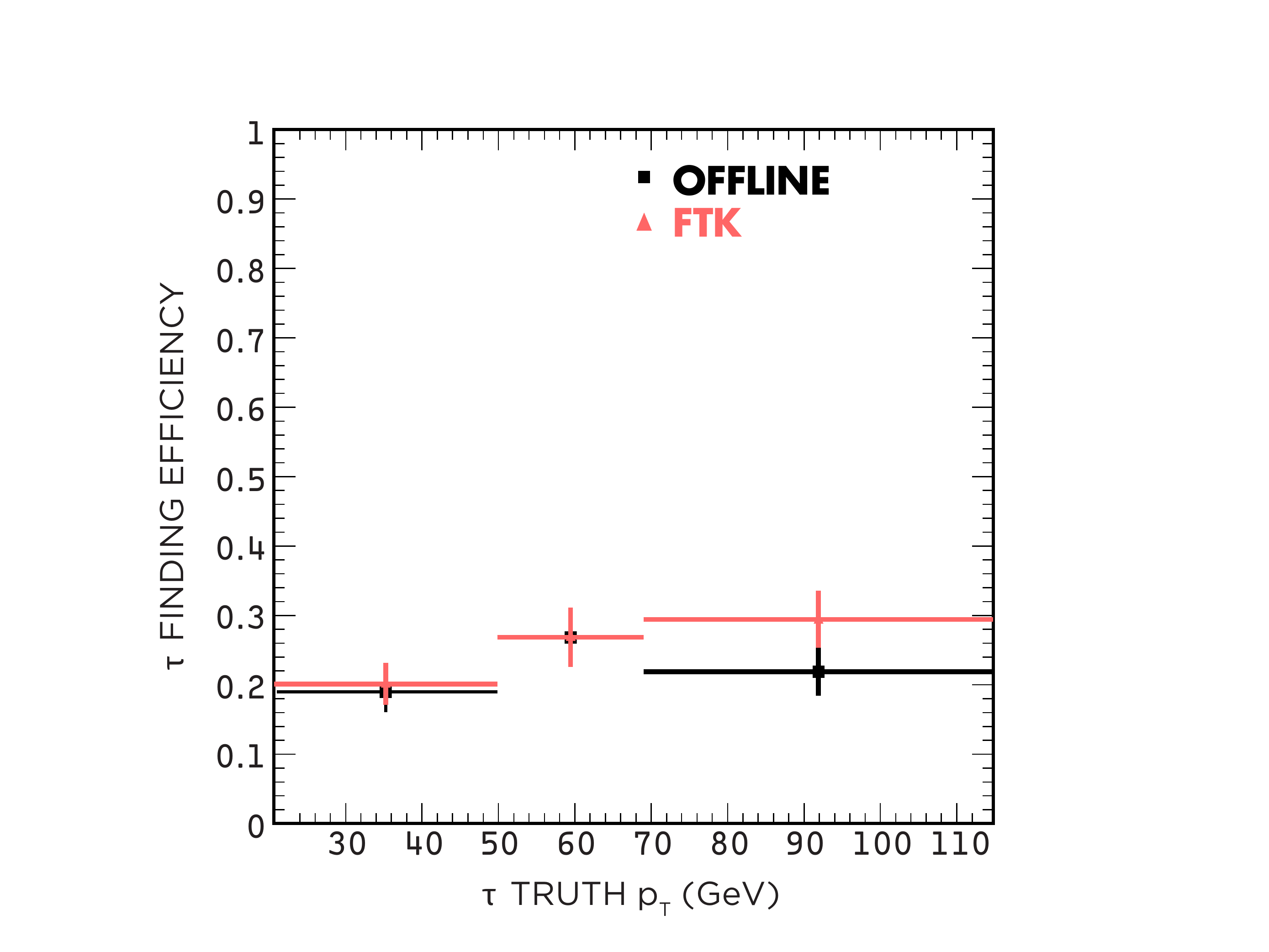}
\caption{$\tau$-tagging efficiency versus $\tau$ $p_T$ for
  $H\rightarrow\tau\tau$ events simulated at $3 \times 10^{34}$
  cm$^{-2}$ s$^{-1}$ for (left) 1-prong and (right) 3-prong hadronic
  $\tau$ decays.}
\label{perf_tautag} 
\end{figure}

As previously mentioned, $b$ quarks and $\tau$ leptons produce specific
tracking signatures which can be exploited in triggering on these
physics objects. Offline-quality $b$-tagging efficiency and light
quark rejection using the signed impact parameter significance can be
achieved by using the savings in tracking time to apply more
sophisticated $b$-tagging algorithms at LVL2. Fig.~\ref{perf_btag}
compares the signed impact parameter significance of FTK tracks with
that of offline tracks. For hadronic $\tau$ decays, a tagging
algorithm that requires 1 (2 or 3) tracks in a signal cone  for
1-prong (3-prong) and no tracks with $p_T > 1.5$ GeV in an isolation
cone. A comparison of the $tau$-tagging efficiency for
$H\rightarrow\tau\tau$ events simulated at $3 \times 10^{34}$
cm$^{-2}$ s$^{-1}$ between offline tracking and FTK tracking is shown
in Fig.~\ref{perf_tautag}. Using this algorithm, the fake $\tau$
probability from jets is found to be $\sim0.1$\% for both offline and
FTK tracking. In both Fig.~\ref{perf_btag} and Fig.~\ref{perf_tautag},
it is shown that FTK tracks are expected to provide tagging
performance that is similar with offline tracks at high luminosity.

\begin{figure}[h]
\centering
\includegraphics[width=80mm]{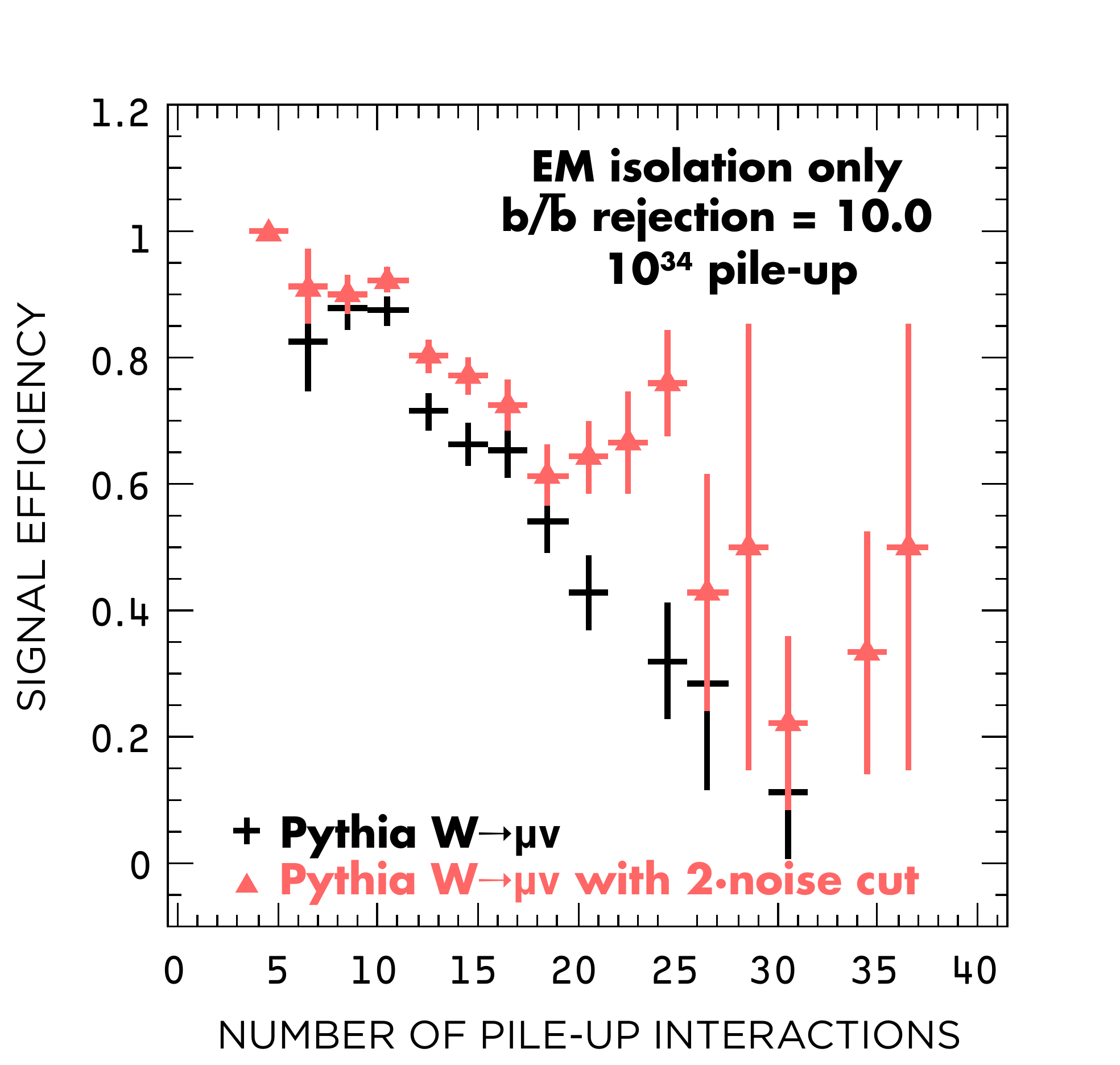}
\includegraphics[width=80mm]{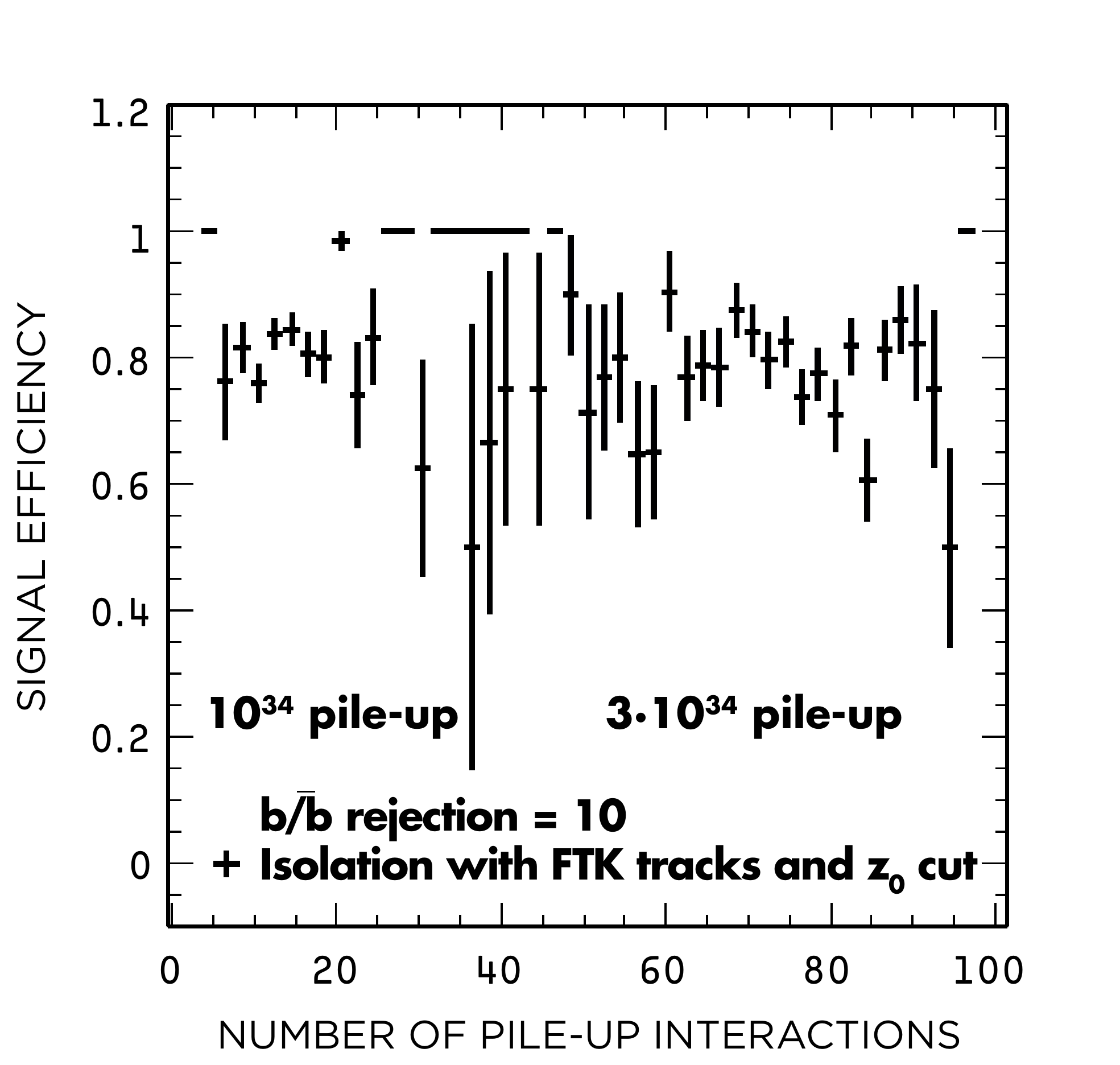}
\caption{Comparison of muon signal efficiency versus the number of
  pile-up interaction using (left) calorimeter-based isolation and
  (right) FTK tracking-based isolation.}
\label{perf_muons}
\end{figure}

Additionally, improved isolation performance facilitated by tracking
information can help retain good efficiency for electron and muon
triggering in the presence of pile-up interactions. A preliminary
study of isolation in muon triggering at high luminosity, where a
purely calorimeter-based isolation becomes less effective, was
performed. The results are summarized in Fig.~\ref{perf_muons}. In
Fig.~\ref{perf_muons} (left), it can be seen that the muon efficiency
using calorimeter-only isolation descreases rapidly with increasing
pile-up, even if cell thresholds are raised on the electromagnetic
calorimeter. Fig.~\ref{perf_muons} (right) shows that a tracking-based
isolation which uses the $z$-position of the vertex has constant muon
signal efficiency out to at least 100 pile-up interactions.

\section{Conclusions and Outlook}
FTK performs global track reconstruction at the full LVL1 trigger
rate and naturally integrates with the current ATLAS data acquisition
system. Using massively parallel Associative Memories, it will provide
a complete list of three-dimensional tracks at the beginning of
LVL2 processing, including tracks outside of the ROIs. The extra
time saved by FTK can be used in LVL2 to apply more advanced
algorithms and ultimately extend the physics reach of the ATLAS
experiment.

FTK robustly and quickly reconstructs tracks at the LHC design
luminosity and produces efficiencies and resolutions comparable to
offline tracking. The FTK execution time for tracking a typical high
luminosity events over all sensitive regions of the detector is
expected to by a 1000 times faster than the current system tracking in
a single ROI. 

The FTK project was approved as an official ATLAS upgrade project in
December 2010. The FTK system is currently under development by groups
in Europe, Japan, and the U.S., with the goal of having a full system
deployed by 2016. Currently, the system architecture and algorithms
are being finalized. Work on designing and building prototype boards
and a next-generation Associative Memory chip is under way.

\end{document}